\documentclass[aps,prl,twocolumn,superscriptaddress,floatfix,noshowpacs,10pt,longbibliography]{revtex4-2}
\usepackage{graphicx,bm,times}
\usepackage{amsmath}
\usepackage{amsfonts}
\usepackage{amssymb}
\usepackage{bm}% bold math
\usepackage{color}
\usepackage{times}
\usepackage[%
  colorlinks=true,
  urlcolor=blue,
  linkcolor=blue,
  citecolor=blue
  ]{hyperref}
\DeclareUnicodeCharacter{2212}{-}
\DeclareUnicodeCharacter{2009}{\,}
\begin{document}

\title{{Robust superconductivity and fragile magnetism induced by \\ the strong Cu impurity scattering  in the high-pressure phase of FeSe}}

\author{Z. Zajicek}
\email[corresponding author:]{zachary.zajicek@physics.ox.ac.uk}
\affiliation{Clarendon Laboratory, Department of Physics, University of Oxford, Parks Road, Oxford OX1 3PU, UK}

\author{S. J. Singh}
\affiliation{Clarendon Laboratory, Department of Physics,
University of Oxford, Parks Road, Oxford OX1 3PU, UK}
\email[corresponding author:]{Current address: Institute of High Pressure Physics (IHPP), Polish Academy of Sciences, Sokolowska 29/37, 01-142, Warsaw, Poland}

\author{A. I. Coldea}
\email[corresponding author:]{amalia.coldea@physics.ox.ac.uk}
\affiliation{Clarendon Laboratory, Department of Physics, University of Oxford, Parks Road, Oxford OX1 3PU, UK}

\begin{abstract}
Superconductivity in FeSe is strongly enhanced under applied pressure and it is proposed to emerge from anomalously coupled structural and magnetic phases.
Small impurities inside the Fe plane can strongly disrupt the pair formation in FeSe at ambient pressure and can also
reveal the interplay between normal  and superconducting  phases.
Here, we investigate how an impurity inside the Fe plane induced by the Cu substitution can alter the balance between competing electronic phases of FeSe at high pressures.
 In the absence of an applied magnetic field, at low pressures the nematic and superconducting phases are suppressed by a similar factor.
On the other hand, at high pressures, above 10~kbar, the superconductivity remains unaltered despite the lack of any signature in transport associated
 to a magnetic phase in zero-magnetic field.
 However, by applying a magnetic field, the resistivity displays an anomaly preceding the activated behaviour in temperature,
 assigned to a magnetic anomaly.  We find that
  the  high-pressure superconducting phase of FeSe is robust and remains enhanced in
 the presence of Cu impurity, whereas the magnetic phase is not.
 This could suggest that high-$T_{\rm c}$
  superconductivity has a sign-preserving order parameter
  in a presence of a rather  glassy magnetic phase.
\end{abstract}

\date{\today}
\maketitle

\paragraph{Introduction.}
Hydrostatic pressure is an invaluable tool to stabilize novel electronic phases as well as to enhance superconductivity towards room temperature \cite{Snider2020}.
Among unconventional superconductors, FeSe displays the signature of a nematic electronic phase
before becoming superconducting at low temperatures below 9\,K \cite{Coldea2017}.
However, with applied pressure the nematic phase is suppressed and
the superconducting transition temperature is enhanced towards 37\,K  close to 6.3\,GPa
 \cite{Mizuguchi2008,Medvedev2009,Garbarino2009,Braithwaite2009,Margadonna2009,Masaki2009,Sun2016pressure}.
This enhanced superconductivity at high pressures occurs in a region in
which a new electronic phase, believed to be of magnetic
origin (spin-density wave (SDW) phase), is present \cite{Bendele2010,Bohmer2019}.
The volume and the local magnetic field of this magnetic phase
 is strongly dependent on the applied pressure \cite{Bendele2010}
and it coincides with a first-order  structural transition at high pressures,
suggesting a  potential magnetoelastic coupling and phase coexistence \cite{Bohmer2019}.
 Interestingly, this magnetic phase
 follows the superconducting phase very closely at high pressures,
and it raises the question whether  bulk superconductivity coexists or
competes on short-length scales with the magnetic order \cite{Bendele2010}.
Furthermore, the enhancement of superconductivity
is also present in FeSe$_{1-x}$S$_x$ under pressure, even though the
signatures associated with any magnetic order are strongly reduced
with increased isoelectronic substitution \cite{Matsuura2017}.

Another tuning parameter of superconducting and nematic phases of FeSe is the chemical substitution either inside or outside the conducting Fe plane.
The Cu substitution is highly disruptive, due to the larger size of the Cu relative to the Fe ions inside the conducting planes,
which introduces significant impurity scattering.
The nematic and superconducting phases are suppressed linearly up to 3\% Cu substitution,
whereas the carrier mobilities are quickly reduced even with a small amount of Cu substitution
\cite{Gong2021,Zajicek2022,Williams2009}.
By increasing the Cu substitution the system undergoes a metal-to-insulator transition
which could lead to the stabilization of  local magnetic moments at Fe sites \cite{Williams2009,Huh2021,Young2010}.
However, under applied hydrostatic pressure the superconductivity of Fe$_{1-x}$Cu$_x$Se
could be restored by suppressing the insulating behaviour in powder samples \cite{Schoop2011}
or by quenching from high pressures  in single crystals \cite{Deng2021}.
 These reports highlight that the high-pressure superconductivity stabilized
under pressure is robust while signatures of any magnetic order are not detected. 
Theoretical studies suggest that non-magnetic disorder can have the potential to enhance
superconductivity in a multi-band system \cite{Gastiasoro2018},
and thus the Cu substitution can be used to assess this proposal and to understand
its unusual manifestation.

In this paper, we report a transport study of the effect of a small amount of
Cu  substitution ($x \sim $ 0.0025) in FeSe under
applied hydrostatic pressure up to 20\,kbar to understand the response of the superconducting, nematic and magnetic phases.
Despite the suppression of the nematic and superconducting phases at low pressures compared to FeSe,
the high pressure superconducting phase is robust against the Cu substitution.
In the absence of an applied magnetic field, we observe no anomaly in resistivity, often associated to a SDW phase in FeSe,
but a broad superconducting transition invoking that the electronic conduction at high pressures above 10\,kbar is still affected by the Cu substitution.
However, in strong magnetic fields we detect an upturn in resistivity which is strongly hysteretic
suggestive of glassy behaviour.
Furthermore, we detect a reduction of the charge carrier mobilities and increased scattering rate
with increasing pressure, in this region remains with robust superconductivity.

\paragraph{Methods.}
Single crystals of Fe$_{1-x}$Cu$_x$Se, with the nominal composition of $x$ = 0.0025(4)
were grown using the KCl/AlCl$_3$ chemical vapour transport method \cite{Chareev2013,Bohmer2016FeSe}, as reported previously in \cite{Zajicek2022}.
 Magnetotransport studies under pressure were carried out in a 16~T Quantum Design PPMS.
The in-plane resistivity $\rho_{xx}$ and Hall $\rho_{xy}$ components were measured using a low-frequency five-probe technique
and were separated by (anti)symmetrizing data measured in positive and negative magnetic
fields with the magnetic field applied along the $c$-axis  ($I ||$($ab$)).
Good electrical contacts were achieved by In soldering  and currents up to 1\,mA (peak-to-peak) were used to avoid heating.
 The pressure studies were performed using a piston-cylinder cell and Daphne Oil 7373  up to 21\,kbar. The pressure at low temperatures was determined via the superconducting transition temperature of Sn after cancelling the remnant magnetic field \cite{Eiling1981}.
Errors in the exact contact positions and size result in errors of up to 13\% of  absolute values of resistivity.
The superconducting transitions, $T_{\rm c}$ and $H_{\rm c2}$, are defined as the offset temperature or field, respectively, unless stated otherwise.
The nematic structural transition, $T_{\rm s}$, and the high pressure magnetic phase transition, $T_{\rm m}$, are
defined by the minimum in the derivative of the resistivity as a function of temperature,
as in previous studies \cite{GuanYu2019}.

\paragraph{Transport studies under applied pressure.}
Fig.~\ref{Fig1}(a) shows the temperature dependence of the resistivity in the absence of a
magnetic field for pressures up to 20\,kbar (see shifted curves in Fig.~S1 in the Supplemental Material (SM) \cite{SM}).
Due to the presence of Cu, the nematic transition $T_{\rm s}$, at 80\,K,
and superconducting transition, $T_c$, at 6.2\,K are suppressed, as compared with FeSe
\cite{Zajicek2022}.
Under applied pressure, the nematic transition is linearly suppressed with applied pressure up to 15\,kbar,
similar to FeSe \cite{Gati2019}, whereas the superconductivity is continuously  enhanced with applied pressure up to 20\,kbar,
without showing any dome like-feature visible for FeSe \cite{Gati2019}.
The superconducting transition width, $\Delta T_{\rm c}$, is around 1.3(2)\,K, before
 increasing by a factor of 2 at higher pressures above 10\,kbar, as shown in Fig.~\ref{Fig1}(b).
In FeSe under pressure the superconducting transition width is sharp ($<$1\,K) at low pressures
before increasing significantly towards 7\,K for pressures
where the magnetic phase is present \cite{GuanYu2019}.
This suggests that either the high pressure superconducting phase becomes inhomogeneous in the presence of another competing phase,
or that strong superconducting and/or magnetic fluctuations are present in this high pressure regime \cite{Gati2019}.

\begin{figure}[htbp]
	\centering
	\includegraphics[trim={0cm 0cm 0cm 0cm}, width=0.9\linewidth,clip=true]{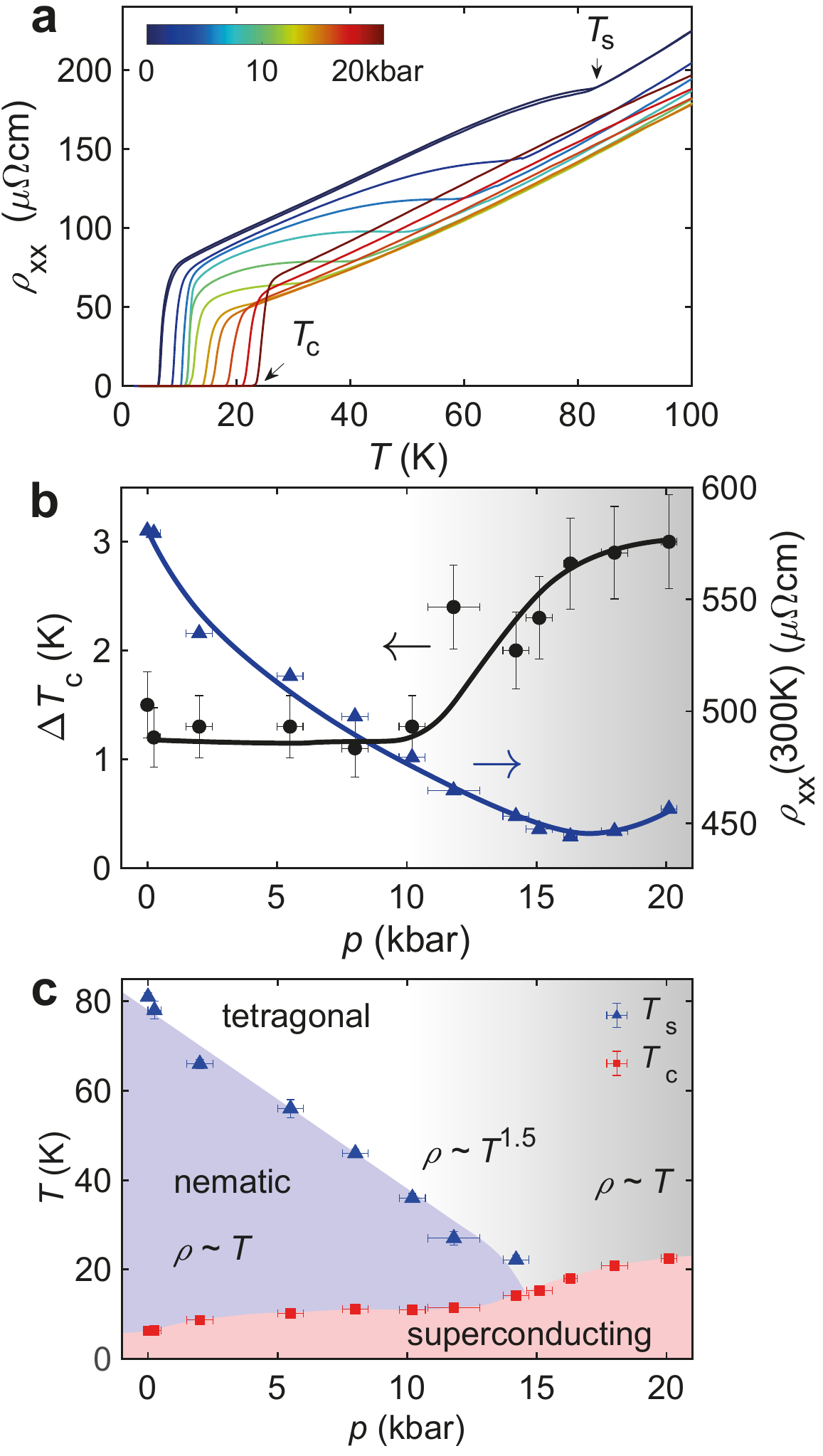}
	\caption{{\bf Zero-field transport properties of Fe$_{1-x}$Cu$_x$Se under pressure.}
		(a) Temperature dependence of the resistivity as a function of applied pressure.
		(b) The evolution of the superconducting transition width, $\Delta T_{c} = T_{\rm on} - T_{\rm off}$ (black circles),
 and the resistivity at 300\,K (blue triangles) with pressure. Solid lines are guides to the eye.
 		(c) Temperature-pressure phase diagram in zero magnetic field tuned by pressure.
The nematic phase occurs at $T_{\rm s}$ (blue triangles)
and superconductivity has the offset temperature at $T_{\rm c}$  (red squares).
The phase diagram can be divided into  three different
regions: the low pressure region up to $p_{1} = 10$\,kbar,
 inside the nematic phase, the high-pressure region above $p_{2} = 15$\,kbar,
 once the nematic phase is fully suppressed,  and the intermediate pressure region between $p_{1}$ and $p_{2}$.
The shaded areas indicate different phase boundaries.
		}
	\label{Fig1}
\end{figure}

The resistivity behaviour provides additional information about the normal electronic phases as a function of pressure.
At room temperature, the resistivity initially decreases with increasing pressure, indicating that
the in-plane transfer integrals are larger under pressure which increases the bandwidth and the system becomes a better metal, as shown in Fig.~\ref{Fig1}(b).
However, at higher pressures above 15\,kbar this trend reverses with resistivity
increasing again, also observed at low temperatures
(see Fig.~S1(b) in the SM \cite{SM}).
% (see Fig.~\ref{SMFig_nvT}b).
The substitution of Cu in FeSe increases the impurity scattering and the residual resistivity,
leading to the suppression of superconductivity and inducing a linear dependence down to 0.4\,K  \cite{Zajicek2022}.
Under pressure, the linear dependence is found both
 inside the nematic phase at low pressures as well as in the high pressure phase,
 as also illustrated by the temperature dependence of the resistivity exponent, $n$ in
 Figs.~S1(c) and (d) in the SM \cite{SM}.
 %Fig.~\ref{SMFig_nvT}a.
At high temperatures in the tetragonal phase, the resistivity can be described by a power law of $T^{1.5}$
which is  rather similar to that observed for both FeSe$_{1-x}$S$_{x}$,
tuned either by the isoelectronic substitution or applied pressure \cite{Bristow2020,Reiss2020}.
Interestingly, in the high pressure regime above 10\,kbar the
resistivity of Cu-substituted FeSe shows no additional anomaly associated  to other phase transitions
 (see Figs.\ref{Fig1}(a) and S1 in the SM \cite{SM}).
This is in contrast to FeSe where a clear upturn in the resistivity anomalies
was associated with the presence of a magnetic transition at a temperature above $T_{\rm c}$
 \cite{Sun2016pressure,Terashima2015,Gati2019,GuanYu2019}.

\begin{figure*}[htbp]
	\centering
	\includegraphics[trim={0cm 0cm 0cm 0cm}, width=0.9\linewidth,clip=true]{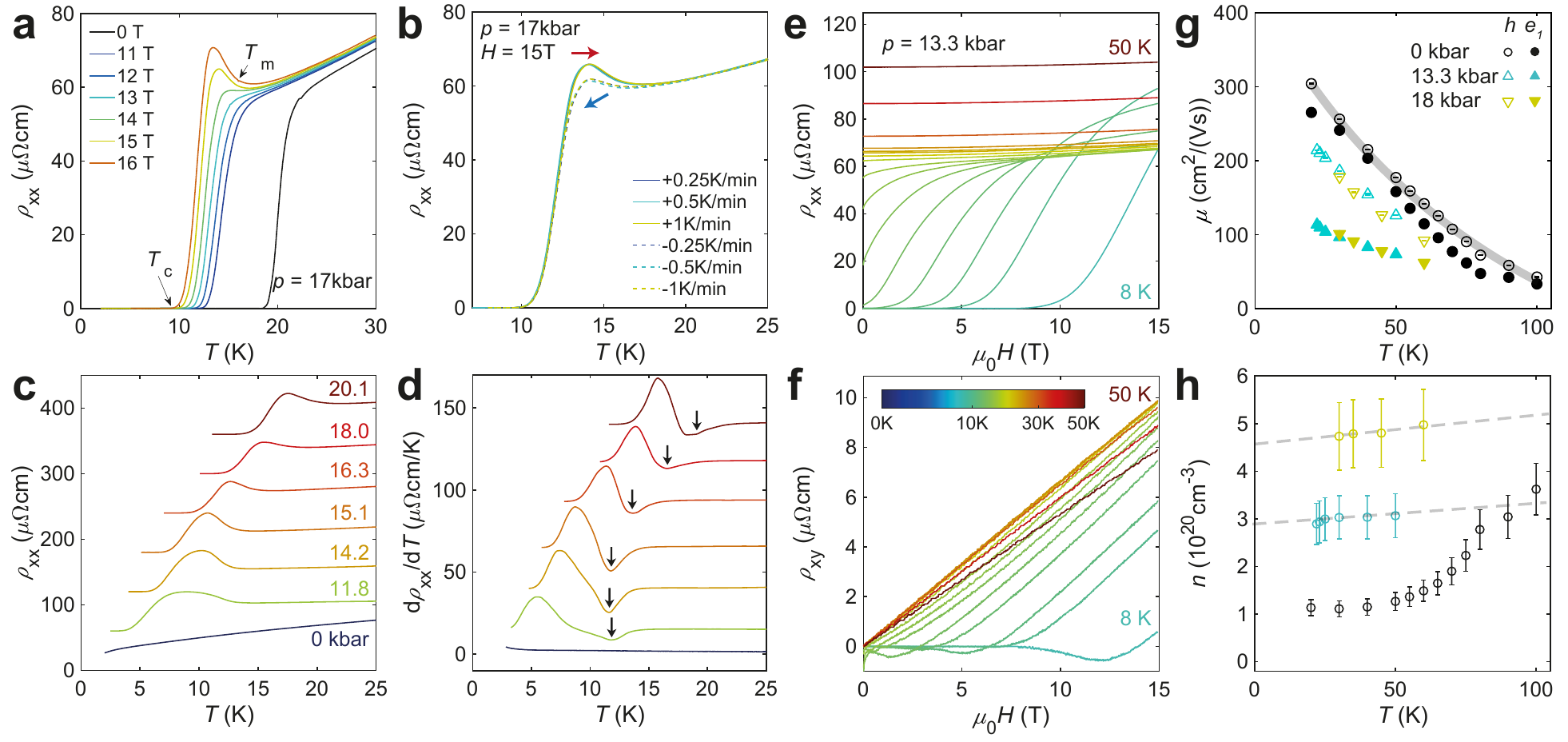}
	\caption{{\bf The transport behaviour in magnetic field of the Fe$_{1-x}$Cu$_x$Se under pressure.}
	(a) The temperature dependence of the
longitudinal resistivity in different magnetic fields at $p$ = 17\,kbar.
	(b) The resistivity upturn at $T_{\rm m}$ in 15\,T at 17\,kbar using various temperature sweep rates.
The solid lines correspond to the warming sweeps whereas the dashed lines correspond to the cooling sweeps.
	(c) The temperature dependence of resistivity in a magnetic field of 15\,T ($H||c$) for different applied pressures
and (d) their corresponding  derivatives. The value of $T_{\rm m}$
is defined as the minimum in the derivative and indicated by arrows, similar to FeSe \cite{Matsuura2017}.
 Curves are offset for clarity.
 Field dependence of (e) the longitudinal resistivity, $\rho_{\rm xx}$, and (f) the Hall resistivity, $\rho_{\rm xy}$ at a pressure of $p$ = 13.3\,kbar.
	(g) Mobility and (h) carrier density of each charge carrier (hole, $h$ and  electron $e_1$)  considering a
  two-band compensated model ($n_h=n_{e_1}$) at different pressures.
  The two-band model at ambient pressure uses data below $<7$~T and a three-band could be used to described
the full field window, as reported in Ref.~\cite{Zajicek2022} and shown in Fig.~S7 in the SM.
Solid and dashed lines are guides to the eye.}
	\label{Fig2}
\end{figure*}

Based on these transport studies, the zero-field temperature-pressure phase diagram
of the Fe$_{1-x}$Cu$_x$Se up to 20\,kbar is constructed, as shown
 in Fig.~\ref{Fig1}(c).
 At high pressures, the superconducting transition temperature, $T_{\rm c}$,
  increases  once the nematicity is suppressed  even in the presence of the Cu impurity.
This behaviour is in agreement with previous studies
in powder samples of Cu substituted FeSe,
where an insulating phase is transformed into a superconductor with applied pressure, although not to a zero resistance state \cite{Schoop2011}.
Furthermore, the phase diagram in zero-magnetic field
in the presence of the Cu impurity inside the conducting plane
 is remarkably similar to  that of  FeSe$_{1-x}$S$_x$,
 in which the isoelectronic substitution takes place outside the conducting plane
 \cite{Reiss2020,Matsuura2017,Xiang2017,Rana2020}.

\paragraph{The high-pressure electronic phase in high magnetic fields.}
Next, we use high magnetic fields to suppress the superconductivity to reveal any hidden electronic phases
and to explore the normal electronic behaviour at lower temperatures, once the superconductivity is suppressed.
Fig.~\ref{Fig2}(a) shows the temperature dependence of resistivity in different magnetic fields at 17\,kbar, in which an
upturn occurs at $T_{\rm m}$, as the temperature is reduced in large enough magnetic field.
This anomaly shows hysteretic behaviour between the cooling and warming curves, independent of the temperature sweep rate,
indicative of a first-order phase transition, as shown in Fig.~\ref{Fig2}(b).
In FeSe, the high pressure magnetic phase displays evidence of hysteresis in heat capacity and NMR studies,
implying the existence of a concomitant structural transition via a spin-lattice coupling \cite{Gati2019,Wang2016pressure}.
The temperature dependence of the resistivity in 15\,T shows smooth changes in the upturn anomaly as a function of the applied pressure, as shown in Fig.~\ref{Fig2}(c) and the corresponding derivatives in Fig.~\ref{Fig2}(d).
As the pressure increases, this anomaly shifts to higher temperatures and it tracks closely the zero field superconducting transition temperature.
In FeSe, this anomaly was associated to a magnetic phase and it occurs already in zero-field
at a higher transition temperature than the superconducting transition, $T_{\rm c}$,
but its appearance seems to vary strongly with pressure, either as an increase, decrease or change in slope \cite{Sun2016pressure}.

\paragraph{Magnetotransport behaviour.}
The magnetotransport studies can provide valuable insight into the changes in the electronic structure and scattering with applied pressure.
Figs.~\ref{Fig2}(e) and (f) show the field dependence of the longitudinal and Hall resistivity, $\rho_{xx}$ and  $\rho_{xy}$, at 13.3~kbar.
The presence of the linear field-dependence of the Hall component indicates a two-band behaviour
which describe the behaviour inside the high pressure and the tetragonal phase (see Figs.~S3, S4, and S5 in the SM \cite{SM}).
%as shown in Fig.~\ref{SMFig_MR_pressure}
%and, as shown Fig.~\ref{SMFig_SS13_ZP1_rho_xx_xy}.
 %\ref{SMFig_MobilityExtras}f
At high pressures, the Hall coefficient extracted from the slope of the Hall resistivity in low fields, $\rho_{xy}$, is positive and generally increases
with decreasing temperature, as shown in Fig.~S7(f) in the SM \cite{SM}.
This is in stark contrast to FeSe which displays negative Hall coefficient
at low temperatures inside the nematic phase both at ambient and at low pressures \cite{Watson2015b,Farrar2022,Sun2017}.

Magnetotransport measurements enable the extraction of the charge carrier density and mobilities
by simultaneously fitting the $\rho_{xx}$ and $\rho_{xy}$ components to a two-band compensated model.
We use the initial input parameters from the mobility spectrum, whose peaks
indicate that the mobilities are slightly reduced with increasing pressure (see Fig.~S6 in the SM \cite{SM}).
%(see Fig.~\ref{SMFig_mobilitySpectrums}.)
In order to compare the parameters
between different pressure regions,
we use a two-band compensated model in low-field regime ($<7$~T)  for ambient pressure
rather the three-band compensated model which can account for the non-linear Hall resistivity, similar to FeSe \cite{Zajicek2022,Watson2015c,Farrar2022}.
Figs.~\ref{Fig2}(g) and (h) compare the temperature dependence of the charge carrier mobilities and carrier densities at various pressures (see also Fig.~S7 in the SM \cite{SM}).
Upon entering the nematic phase from high temperatures there is a significant decrease
in the apparent  carrier density, $n$, due to the development of anisotropic scattering, as found for FeSe \cite{Watson2015c,Farrar2022}.
In the high pressure phase,
the carrier density  is significantly larger indicative of an increase of the Fermi surface size
similar to findings for FeSe$_{1-x}$S$_x$  \cite{Reiss2020}.
Furthermore, the mobilities of both holes and electrons are suppressed with applied pressure,
with stronger suppression
for the negative charge carriers, similar to findings in thin flakes of FeSe \cite{Farrar2022}.
This could indicate an increased scattering rate of spin fluctuations at high pressures and/or an increase in effective mass,  which is consistent
with the enhanced resistivity found in the high pressure phase of Cu-substituted FeSe shown in Fig.~\ref{Fig1}(b).
This behaviour is rather similar to that of FeSe under pressure,
where a three-band model is required to explain its low temperature behaviour for all pressures,
but the carrier densities are enhanced and the mobilities decreased with increasing pressure
\cite{Terashima2016_MR}.

\paragraph{The pressure-temperature $p-T$ phase diagrams.}
To summarize we compare
the temperature-pressure phase diagram of Cu-FeSe
with that of FeSe \cite{GuanYu2019} in 0~T and 15~T, as shown in Figs.~\ref{Fig3}(a) and (b).
The phase diagram can be split into three distinct regions: firstly, in the low pressure region,
$p < p_{1} \sim 10$\,kbar, the superconducting phase emerges from the nematic phase,
secondly at intermediate pressures, $p_{1} < p < p_{2} \sim 15$\,kbar,  signatures of the magnetic phase are detected in FeSe in zero-field  and
 finally, at higher pressures, $p > p_{2}$ the magnetic and superconducting phases coexist and the nematic phase is suppressed at $p_{2}$.
Interestingly, as the pressure increases beyond $p_1$, $T_{\rm c}$
displays a broadly similar pressure dependence for the two systems,
indicating a robustness of superconductivity to impurity scattering.
The increase in $T_{\rm c}$
occurs in a regime where the carrier density $n$ increases at high pressure, as shown in Fig.~\ref{Fig2}(h).
On the other hand, the magnetic
phase above $p_2$ is highly sensitive to impurity scattering, being
washed out by the small amount of the Cu substitution and its signatures are only
visible in magnetic field, as shown in Fig.~\ref{Fig3}(b).
The upper critical field across these phases shows a remarkable scaling to $T_{\rm c}$
in the presence of the Cu impurity across the phase diagram (Fig.~S7 in the SM).
% (Fig.~\ref{Fig:Hc2}).

In order to account for the relative changes in superconductivity
due to the Cu substitution,
 the temperature-pressure diagrams are scaled to their ambient value
of $T_{\rm s}$, as shown in Fig.~\ref{Fig3}(c). We notice a remarkable scaling
of the nematic and superconducting temperatures between FeSe and Cu-FeSe,
up to $p_{1}$, suggesting that the impurity scattering affects
the two phases in a similar manner and have a similar origin,
as found also for higher Cu substitution \cite{Zajicek2022,Gong2021}.
As the superconductivity at low pressure is strongly suppressed by the Cu substitution, this is consistent with
sign-changing $s_{\pm}$-type of order parameter in the low pressure regime \cite{Zajicek2022}.

\begin{figure}[htbp]
	\centering
	\includegraphics[trim={0cm 0cm 0cm 0cm}, width=0.75\linewidth,clip=true]{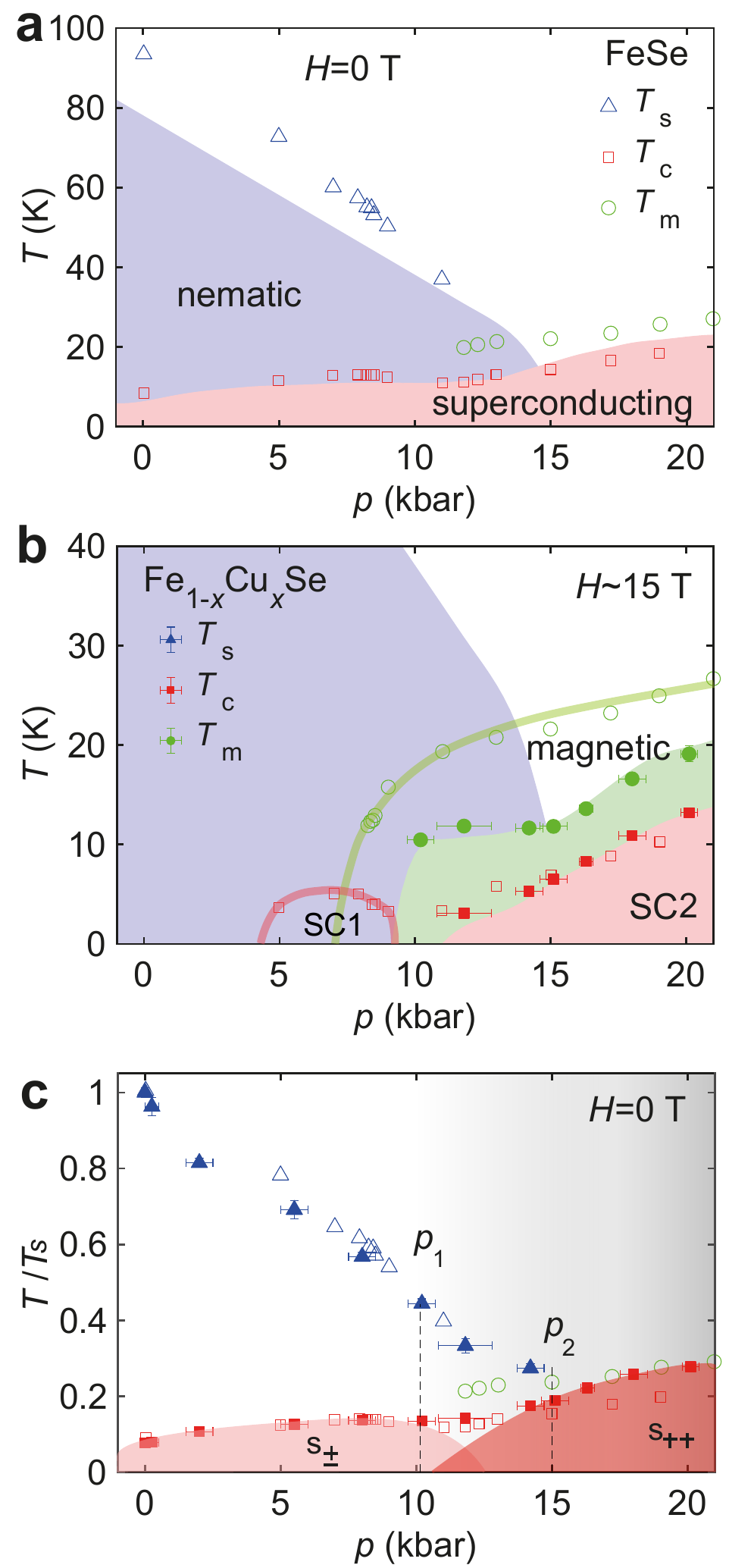}
	\caption{ {\bf The pressure-temperature phase diagrams of Cu substituted FeSe.}
	 The comparison between the different electronic phases of FeSe (open symbols)
from Ref.~\onlinecite{GuanYu2019} and Cu-FeSe (solid symbols)
measured in (a) 0~T and (b) in 15~T for Cu-FeSe and 16~T
for FeSe (after Ref.~\cite{GuanYu2019}).
The shaded areas in all panels reflect the phase boundaries for Cu-FeSe, as
described in Fig.~\ref{Fig1}(c). The solid lines in (b) are guides to the eye for FeSe.
 Nematic shaded area in (a) and (b) are for Cu-FeSe, assuming a field-independent $T_{\rm s}$.
	(c) The scaling of the temperature-pressure phase diagram from  (a)
by using the reduced temperature $t=T/T_{\rm s}$, where $T_{\rm s}$ is the nematic
transition at ambient pressure for the two systems.
 The two shaded regions reflect schematically the two proposed superconducting
regions with the boundaries given by the maximum value of $T_{\rm c}$.
}
	\label{Fig3}
\end{figure}

Interestingly, at high pressure above $p_2$ the magnetic and superconducting
phases of FeSe enhance their transition temperatures
with increasing pressure \cite{Gati2019}.
However, the very small Cu substitution is sufficient to break down this trend and only superconductivity is present, whereas the signatures of the magnetic phase in resistivity have disappeared.
This suggest the fragility of the magnetic phase which is only detected and stabilized by the presence of a magnetic field.
In high magnetic fields ($\sim 15$~T), the  magnetic transition
is strongly suppressed in the Cu-FeSe as compared with FeSe,
whereas the superconducting domes are rather similar, as shown in Fig.~\ref{Fig3}(b). 
This suggest that the superconducting phase at high pressure is robust
in the presence of the Cu impurity.

Significant disruption of the high pressure magnetic phase was also detected under pressure
in FeSe crystals with higher disorder \cite{Okabe2010,Miyoshi2014},
as well as in thin flakes of FeSe under pressure,
 where the magnetic phase is suppressed as the thickness decreases \cite{Xie2021}.
Interestingly, in the presence of sulphur substitution
outside the Fe planes, the magnetic phase is also suppressed at lower pressures in FeSe$_{1-x}$S$_x$,
although it was suggested to be stabilized at higher pressures around 50~kbar \cite{Matsuura2017}.
The robust superconductivity at high pressure occur up to 31~K also in the presence of large
amount of the Cu substitution towards 4\% but there are no signatures of a
magnetic phase \cite{Okabe2010,Miyoshi2014,Schoop2011}.
This implies that the superconductivity in the high pressure phase of FeSe and Cu-FeSe
 is rather robust to impurity scattering,
which would be consistent
 with an $s_{++}$ pairing symmetry or that the high-pressure superconductivity is strongly inhomogeneous.
In the latter case, the superconducting order could still have a $s_{\pm}$ pairing symmetry
 since extended emergent disorder regions
are unable to provide large-momentum transfer interband scattering
connecting opposite signs of the superconducting gap function on electron and hole pockets,
and therefore strongly limiting any substantial pair-breaking processes from the inhomogeneous disorder landscape.
However, for superconductors with a sign-reversal order parameter
the $T_{\rm c}$ is strongly suppressed by impurity scattering, as found for
the Cu-substituted FeSe inside the nematic phase \cite{Zajicek2022}.
Theoretically, it is suggested that disorder in multi-band superconductors
disorder can cause a transition between $s_{+-}$ and $s_{++}$ pairing symmetry
if one of two gaps changes sign
 %when going through zero
 and $T_{\rm c}$ is predicted to remain finite and almost
independent of the impurity scattering rate \cite{Korshunov2016}.
Additionally, superconductivity could
 be enhanced for a sign-preserving gap, even in the presence of large enough disorder,
as disorder can induce spatial modulations where the coherence
length is a few nanometers \cite{Gastiasoro2018,Efremov2011}.
Fig.~\ref{Fig3}(c) indicates the separation into
 two superconducting domes of potentially different symmetry and response to impurity scattering
of the $p-T$  phase diagram of Cu-FeSe.
Interestingly, two superconducting domes are also present
for FeSe$_{0.89}$S$_{0.11}$,
 in which the isoelectronic substitution occurs outside the conducting plane,
 and no signature of a magnetic phase
is detected at the high-pressure phase up to 22~kbar~\cite{Reiss2020}.

%\paragraph{Discussion.}
The enhanced resistivity at high pressures, combined with the reduction of mobilities while the apparent carrier density increases,
imply that either the high-pressure phase is highly inhomogeneous
or that there are very strong spin fluctuations that do not permit the stabilization of the long-range magnetic order.
These effects are also evident in the evolution of the superconducting transition width
which becomes extremely broad  above $p_{1}$ both in FeSe and Cu-FeSe, as shown in Fig.~\ref{Fig1}b.
The large hysteresis effects observed in the resistivity curves for different cooling rates in the high pressure phase
suggest that the local short-range magnetic order is affected by the presence of Cu ions, by freezing of dynamical spin fluctuations of the Fe magnetic moments in the presence of non-magnetic disorder \cite{Andersen2010}.
Evidence of  short-range local magnetism was found in FeSe under pressure
using $\mu$SR as well as Mossbauer spectroscopy   \cite{Bendele2010,Bohmer2019},
 but neutron diffraction studies
 have not been able to resolve any long-range magnetic order
  \cite{Bendele2012}.

Theoretical calculations of non-magnetic impurities in FeSe show
 that point-like disorder sites locally induce short-range magnetism \cite{Martiny2019,Gastiasoro2014}.
The presence of the Cu impurity inside the SDW phase
can lead to phase fragmentation or glassy behaviour,
similar to a SDW glass proposed to arise in systems with weak non-magnetic disorder~\cite{Mross2015}.
Microscopic calculations inside the SDW phase of iron-based superconductors
have suggested that a strong non-magnetic potentials, like Cu, can generate extended magnetic nematogens,
which are regions  of a
a competing magnetic phase centered at the impurity site
(typical lengths of $\sim  10$ lattice constants) \cite{Gastiasoro2014}.
Such entities could be  short-circuited in transport studies at high pressure
leading to the vanishing of the SDW signatures, despite the tiny amount of Cu in the samples.
An external field, however, could potentially re-align and couple
the induced magnetic puddles
 and thereby introduce additional spin-dependent scattering
 which can enhance resistivity and mimic the behaviour
 in the absence of the Cu impurities.
Spatially-resolved probes would be very valuable for obtaining
 a detailed understanding of the fascinating interplay between disorder,
 magnetism and superconductivity in the high-pressure region of FeSe.
Theoretical work addressing the nature and the robustness of the high-pressure
superconducting phase, specific to FeSe and its related systems,  would be very
valuable to guide experimental findings and to help
identify the essential ingredients to design a high-temperature superconductor.

\paragraph{Conclusions}

Our study provides new insight into the nature of unconventional
superconductivity of FeSe in different regimes and its sensitivity to impurity
scattering in the presence of a small amount of Cu substitution.
The magnetic phase at high pressures is profoundly affected by the impurity
scattering that can lead to a phase separation between the
superconducting and magnetic phases with a hysteretic first-order transition.
At low pressures, nematic and superconducting phases are suppressed by the Cu impurity,
consistent with  the presence of a sign-changing pairing.
 On the other hand, the robustness of the high-pressure superconducting
 phase, despite the fragility of the magnetic phase,
  implies  either a non-sign changing superconducting phase,
 or a highly inhomogeneous phase with mainly extended scattering
 regions  with suppressed pair-breaking effects for $s_{\pm}$ superconductivity.
Further theoretical work is required to understand how
to stabilize such a robust superconductivity,
in the absence of applied pressure.

\paragraph{Acknowledgements}
We thank Brian Andersen, Peter Hirschfeld, Maria Gastiasoro, and Pascal Reiss for very useful discussions and comments on the manuscript.
 This work was mainly supported by EPSRC (EP/I004475/1) and Oxford Centre for Applied Superconductivity.
 We also acknowledge financial support of the John Fell Fund of the Oxford University.
 Z.Z. acknowledges financial support from
the EPSRC studentship (EP/N509711/1).
AIC acknowledges an EPSRC Career Acceleration Fellowship (EP/I004475/1).

\bibliography{FeCuSe_bib_Feb2022}

\clearpage
\newpage

%%%%%
\title{Supplemental Material}
\maketitle

\newcommand{\blue}{\textcolor{blue}}
\newcommand{\bdm}[1]{\mbox{\boldmath $#1$}}

\renewcommand{\thefigure}{S\arabic{figure}} % changes FIG.~1 to FIG.~SM1
\renewcommand{\thetable}{S\arabic{table}} % changes FIG.~1 to FIG.~SM1

\newlength{\figwidth}
\figwidth=0.48\textwidth

\setcounter{figure}{0}

\newcommand{\fig}[3]
{
	\begin{figure}[!tb]
		\vspace*{-0.1cm}
		\[
		\includegraphics[width=\figwidth]{#1}
		\]
		\vskip -0.2cm
		\caption{\label{#2}
			\small#3
		}
\end{figure}}
%%%%

%\begin{figure}[htbp]
%	\centering
%	\includegraphics[trim={0cm 0cm 0cm 0cm}, width=0.8\linewidth,clip=true]{Fig_SM_rhovT_0T.pdf}
%	\caption{{\bf Zero-field resistivity of Fe$_{1-x}$Cu$_{x}$Se under pressure.}
%Temperature dependence of the resistivity as a function of applied pressure. The curves are shifted for clarity (by 75$\mu\Omega cm$ each). The pressure of each run is given by the corresponding colour. The non-shifted curves are shown in Fig.~1(a).	
%}
%	\label{SMFig_rhovT_Waterfall}
%\end{figure}

\begin{figure*}[htbp]
	\centering
	\includegraphics[trim={0cm 0cm 0cm 0cm}, width=0.8\linewidth,clip=true]{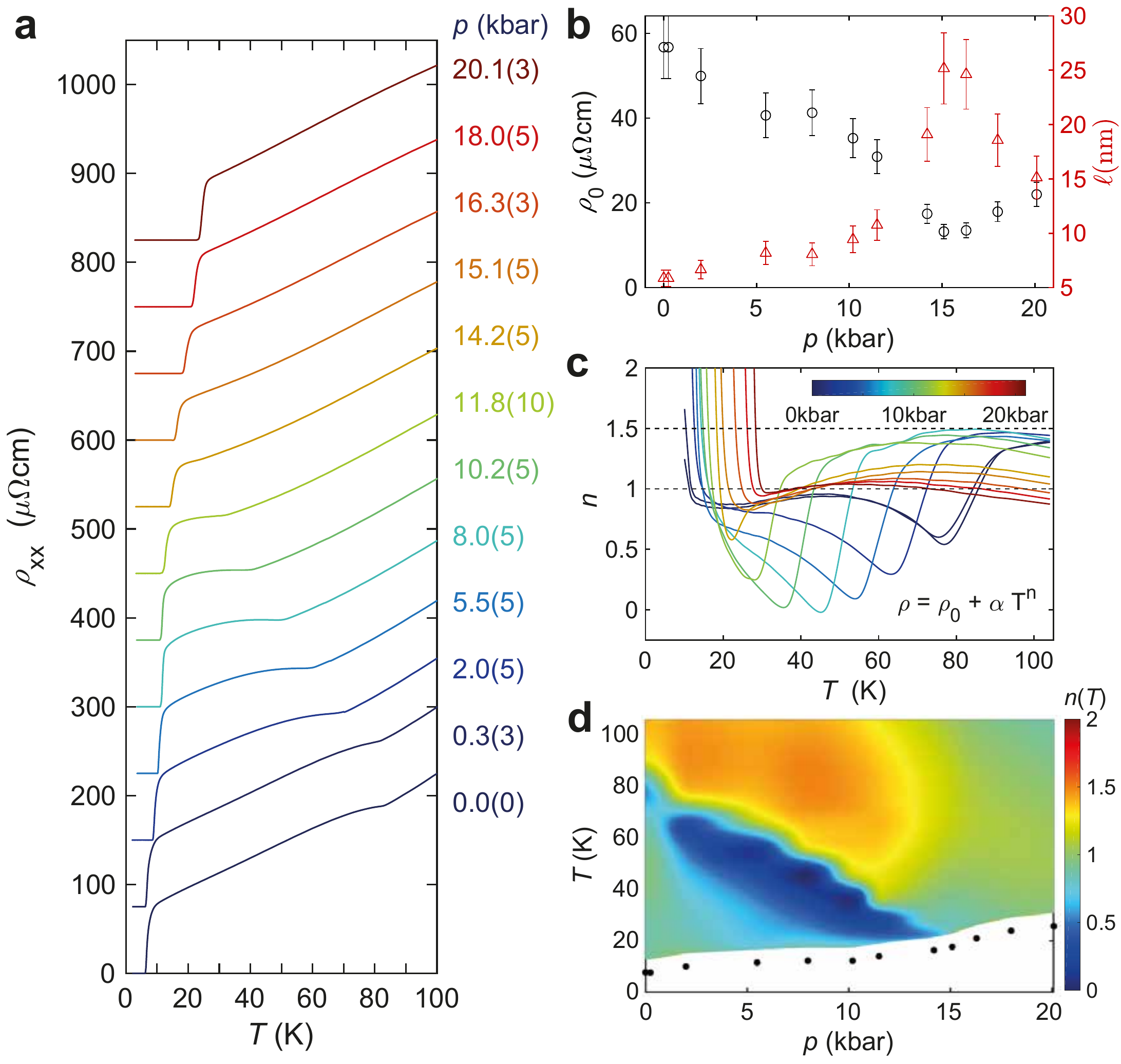}
	\caption{{\bf The zero-field resistivity and the local resistivity exponent.}
(a) Temperature dependence of the resistivity as a function of applied pressure.
The curves are shifted for clarity (by 75$\mu\Omega cm$ each). The pressure of each run is given by the corresponding colour. The non-shifted curves are shown in Fig.~1(a).	
	The resistivity can be expressed using the form $\rho = \rho_{0} + \alpha T^{n}$, where $n$ is the local resistivity temperature exponent.
At ambient pressure the zero-temperature resistivity, $\rho_{0}$, is obtained by using high magnetic fields to suppress
the superconductivity at temperatures below $T_{\rm c}$ and extrapolating the high-field dependence at zero-field for each temperature \cite{Zajicek2022}.
The mean free path, $\ell = \frac{\pi c \hbar}{N e^{2} k_{\rm F} \rho_{\rm 0}}$, is calculated using the ambient
pressure values for $c$ and $k_{\rm F}$ \cite{Zajicek2022}.
%For 0 to 20\,kbar, $c$ for FeSe changes at a rate of -0.0088 \AA/kbar, from a starting value of 5.5237\AA, which is not a significant change.
%Similarly, $k_{\rm F}$ will change with pressure as is known from FeSe quantum oscillation measurements \cite{Terashima2016},
%however the exact change is not known for Cu-FeSe and so is assumed to be constant for this analysis.
%Additionally, it is unknown exactly how large the Fermi surface is for Cu-FeSe inside the magnetic phase.
(b) The pressure dependence of the zero-temperature resistivity, $\rho_{0}$,
extracted from a linear extrapolation of the zero-field resistivity above the onset of superconductivity similar to the ambient pressure dependence \cite{Zajicek2022}.
The red triangles are the mean free path, using the $\rho_{0}$ values.
(c) The temperature dependence of the local resistivity exponent, $n$, for different pressures in zero magnetic field
and (d) its interpolated colour map for different pressures.	
}
	\label{SMFig_nvT}
\end{figure*}

%\begin{figure}[htbp]
%	\centering
%	\includegraphics[trim={0cm 0cm 0cm 0cm}, width=1\linewidth,clip=true]{Figure_HallCoeff.pdf}
%	\caption{{\bf Hall coefficient}
%The Hall coefficient, $R_{\rm H}$, temperature dependence for different pressures of Cu-FeSe. The dashed lines are guide to the eye. The solid line for 13.3\,kbar is from a temperature sweep. T$^{*}$ marks the minimum in $R_{\rm H}$ for 0\,kbar.
%}
%	\label{SMFig_R_H}
%\end{figure}

\begin{figure*}[htbp]
	\centering
	\includegraphics[trim={0cm 0cm 0cm 0cm}, width=1\linewidth,clip=true]{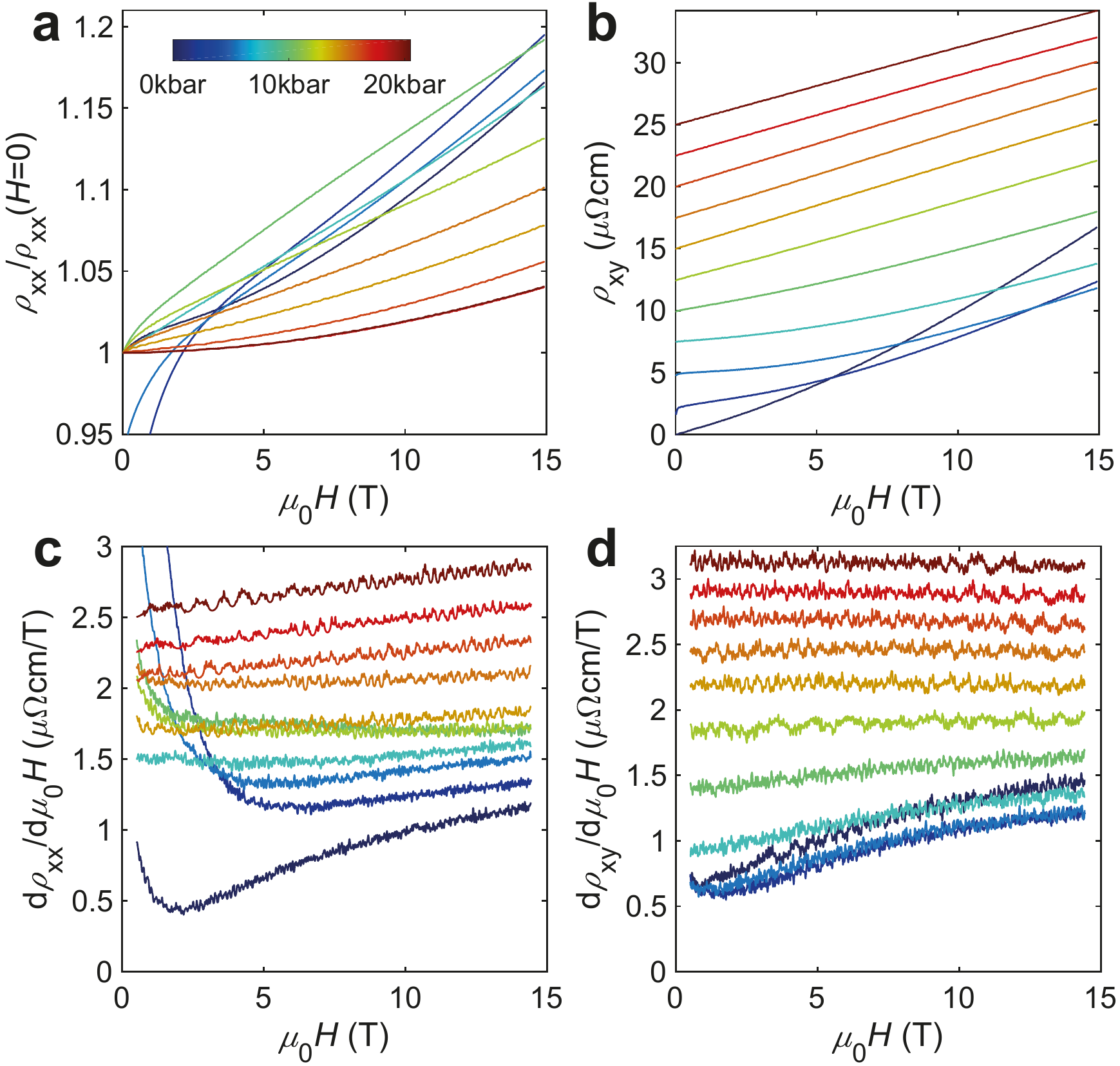}
	\caption{{\bf Magnetotransport behaviour in the normal phase.}
The field dependence of 	(a) $\rho_{xx}$ and (b) $\rho_{xy}$ at a fixed temperature
close to the onset of superconductivity for different pressures
and their corresponding derivatives in relation to the magnetic field in (c) and (d),  respectively.
	The derivatives in magnetic field of the $\rho_{xy}$  component shows a constant behaviour
at high-pressures above 10\,kbar  indicative of a two-band behaviour.}
	\label{SMFig_MR_pressure}
\end{figure*}

\begin{figure*}[htbp]
	\centering
	\includegraphics[trim={0cm 0cm 0cm 0cm}, width=1\linewidth,clip=true]{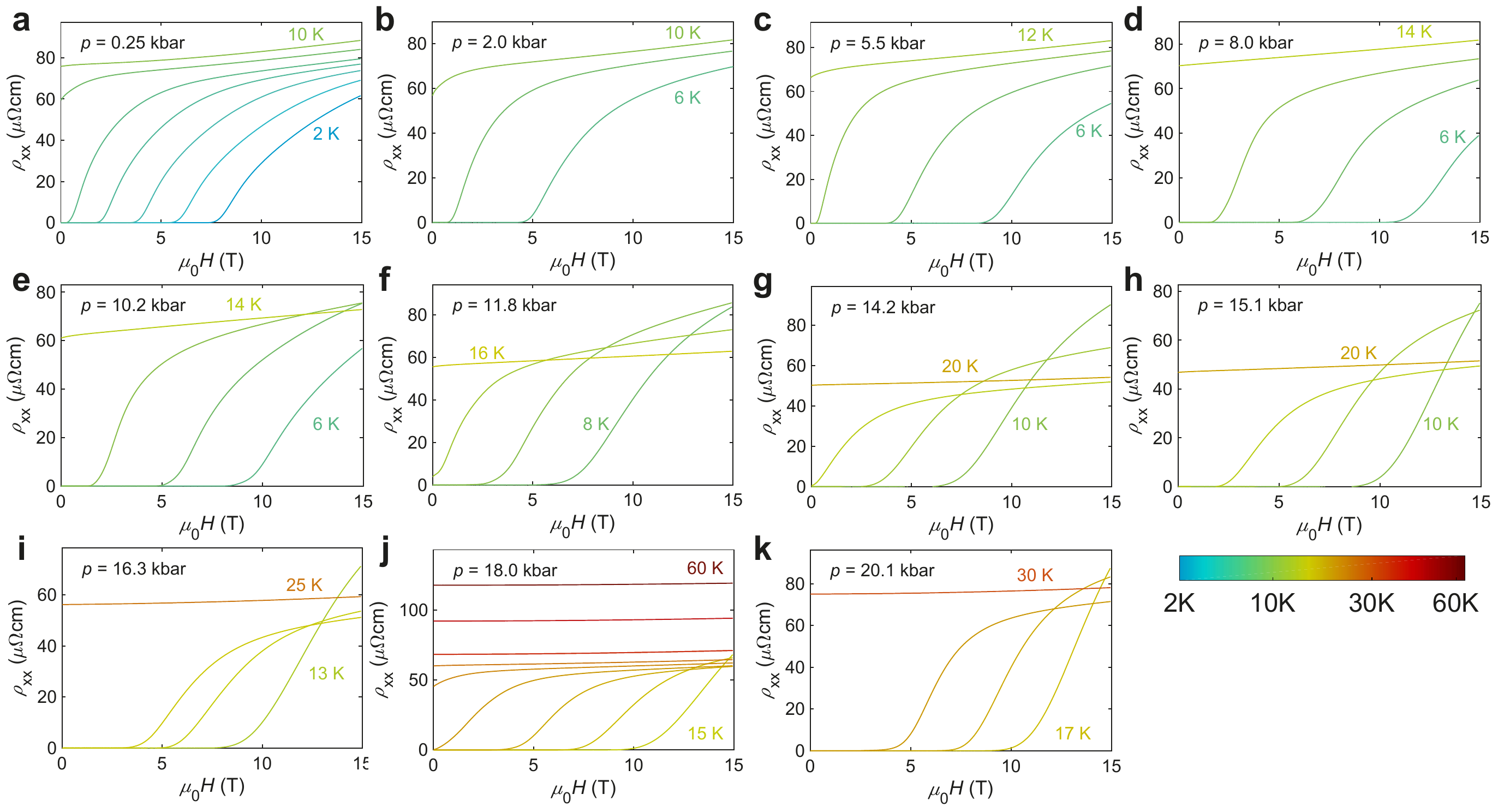}
	\includegraphics[trim={0cm 0cm 0cm 0cm}, width=1\linewidth,clip=true]{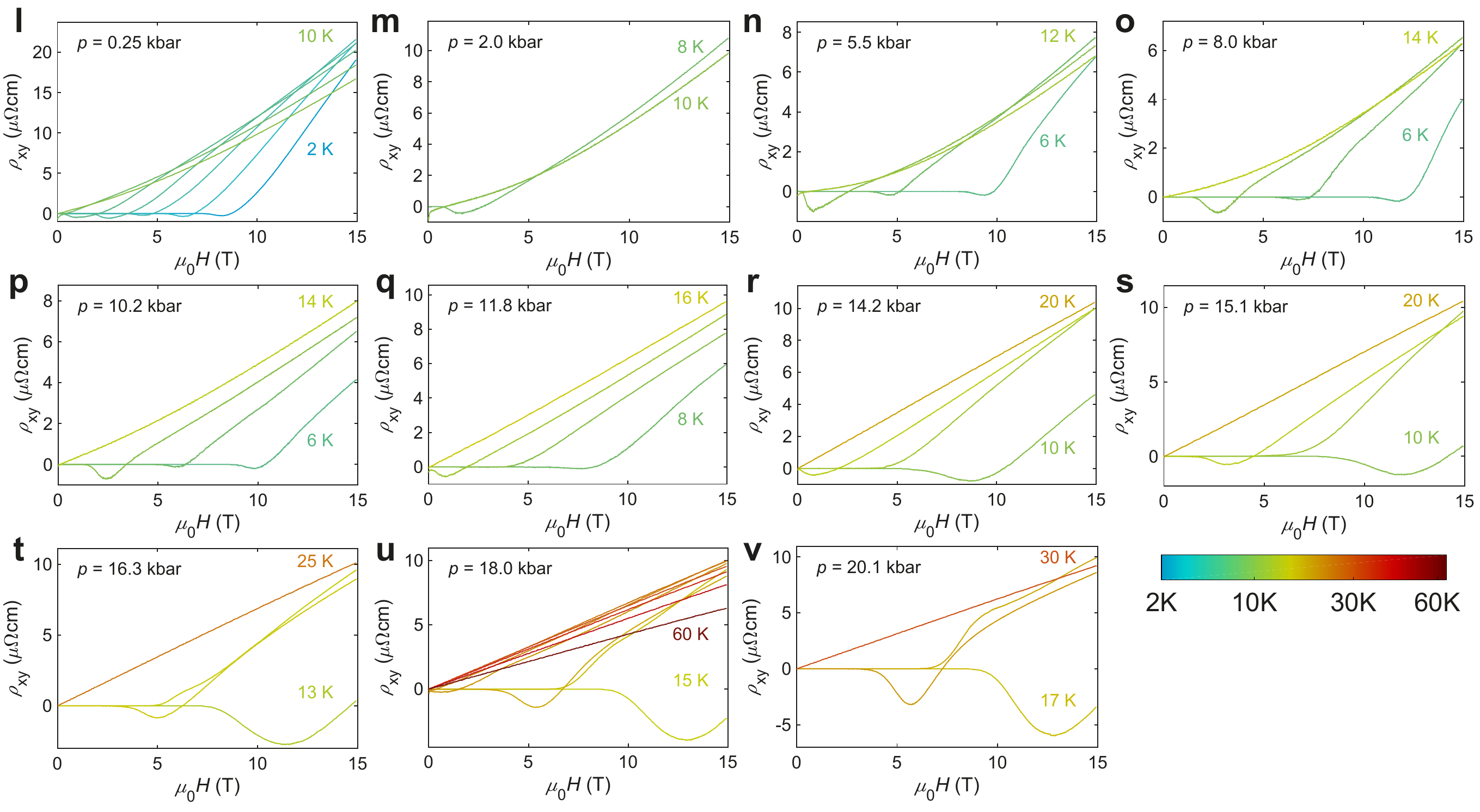}
	\caption{{\bf Magnetotransport behaviour at different pressures.}
The field-dependence of (a) $\rho_{xx}$ and (b) $\rho_{xy}$
measured a different constant temperatures. Each panel corresponds to different applied pressure.}
	\label{SMFig_SS13_ZP1_rho_xx_xy}
\end{figure*}

\begin{figure*}[htbp]
	\centering
	\includegraphics[trim={0cm 0cm 0cm 0cm}, width=1\linewidth,clip=true]{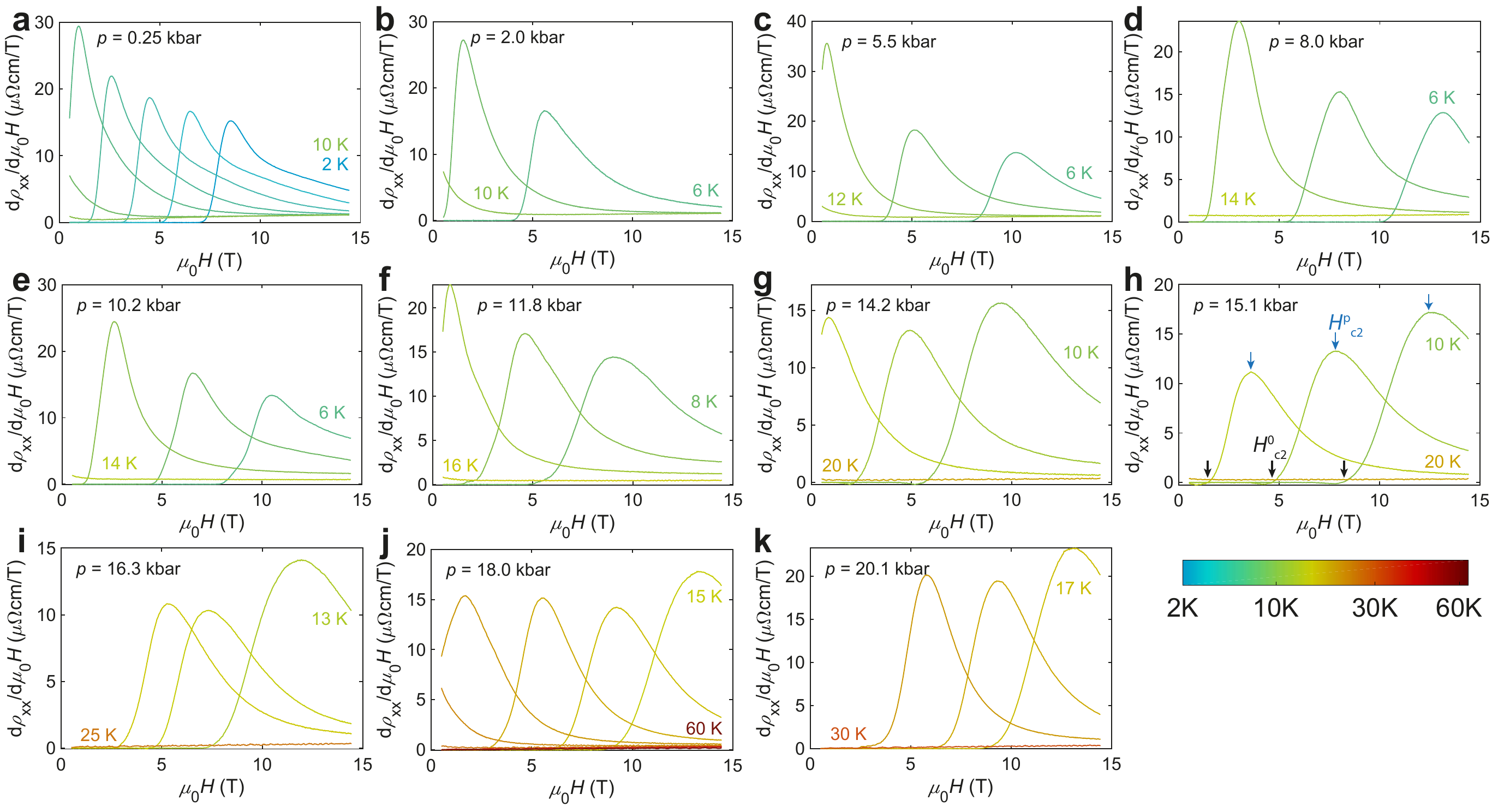}
	\includegraphics[trim={0cm 0cm 0cm 0cm}, width=1\linewidth,clip=true]{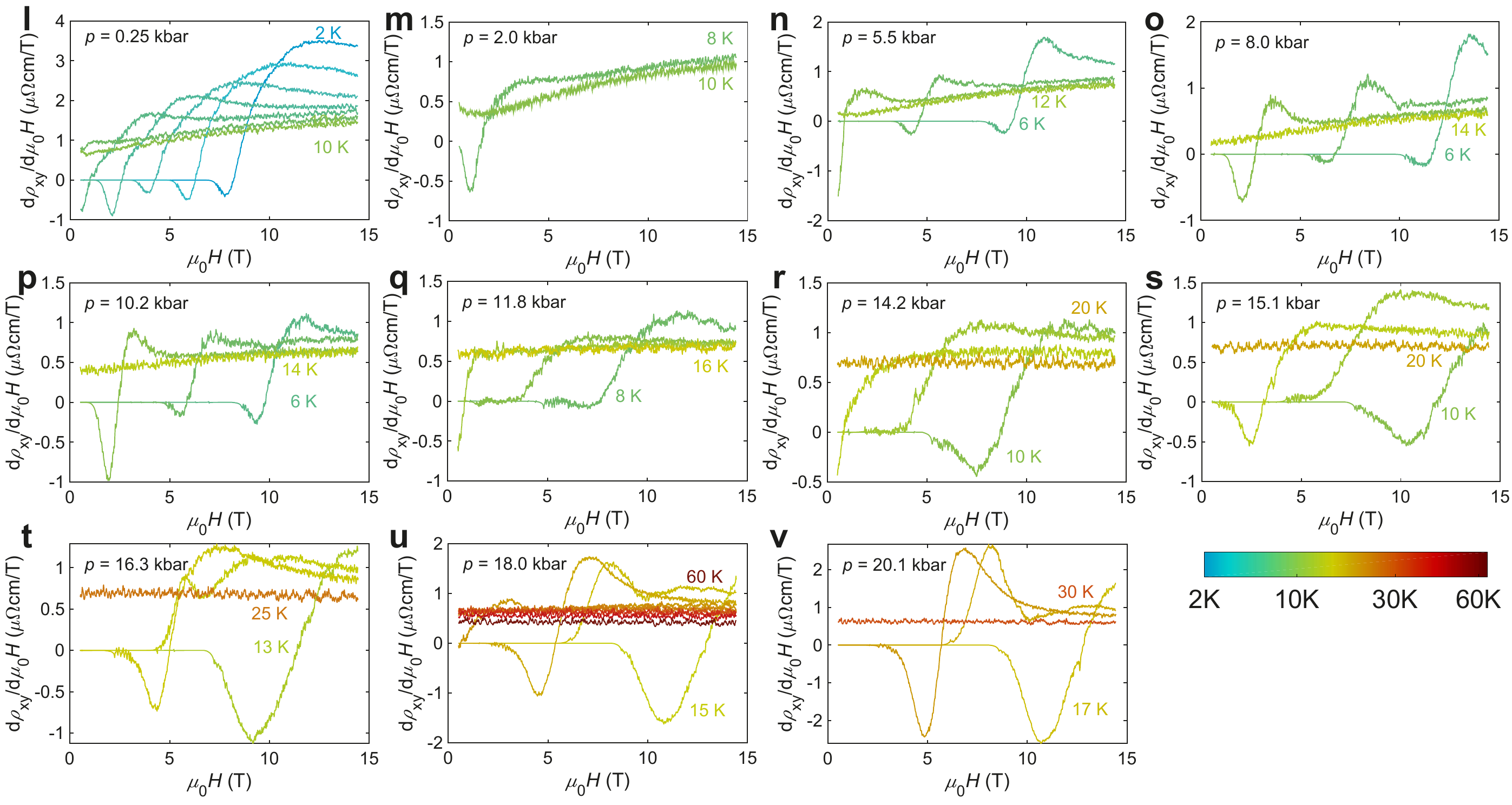}
	\caption{{\bf The derivative of magnetotransport curves in relation to magnetic field at different pressures.}
		Derivatives with respect to magnetic field for
(a) $\rho_{xx}$ and (b) $\rho_{xy}$ as a function of magnetic field measured at different constant temperatures.
Each panel corresponds to different applied pressures.
Examples of how the critical field was extracted are shown in (h). The peak of the derivative is referred to as $H^{\rm p}_{\rm c2}$, and the zero extrapolation as $H^{\rm 0}_{\rm c2}$.
	}
	\label{SMFig_SS13_rhoxx_rhoxy_vB_derivatives}
\end{figure*}

%\begin{figure*}[htbp]
%	\centering
%	\includegraphics[trim={0cm 0cm 0cm 0cm}, width=1\linewidth,clip=true]{FeCuSe_SS13_ZP1_18kbar_rhoXX_rhoXY_v_T.pdf}
%	\caption{{\bf Magnetotransport of FeCuSe SS13 ZP2 under pressure.}
%		(a) $\rho_{xx}$ and (b) $\rho_{xy}$ as a function of magnetic field at constant temperatures.}
%	\label{SMFig_SS13_ZP1_18kbar_rhoXX_rhoXY_v_T}
%\end{figure*}

\begin{figure*}[htbp]
	\centering
	\includegraphics[trim={0cm 0cm 0cm 0cm}, width=1\linewidth,clip=true]{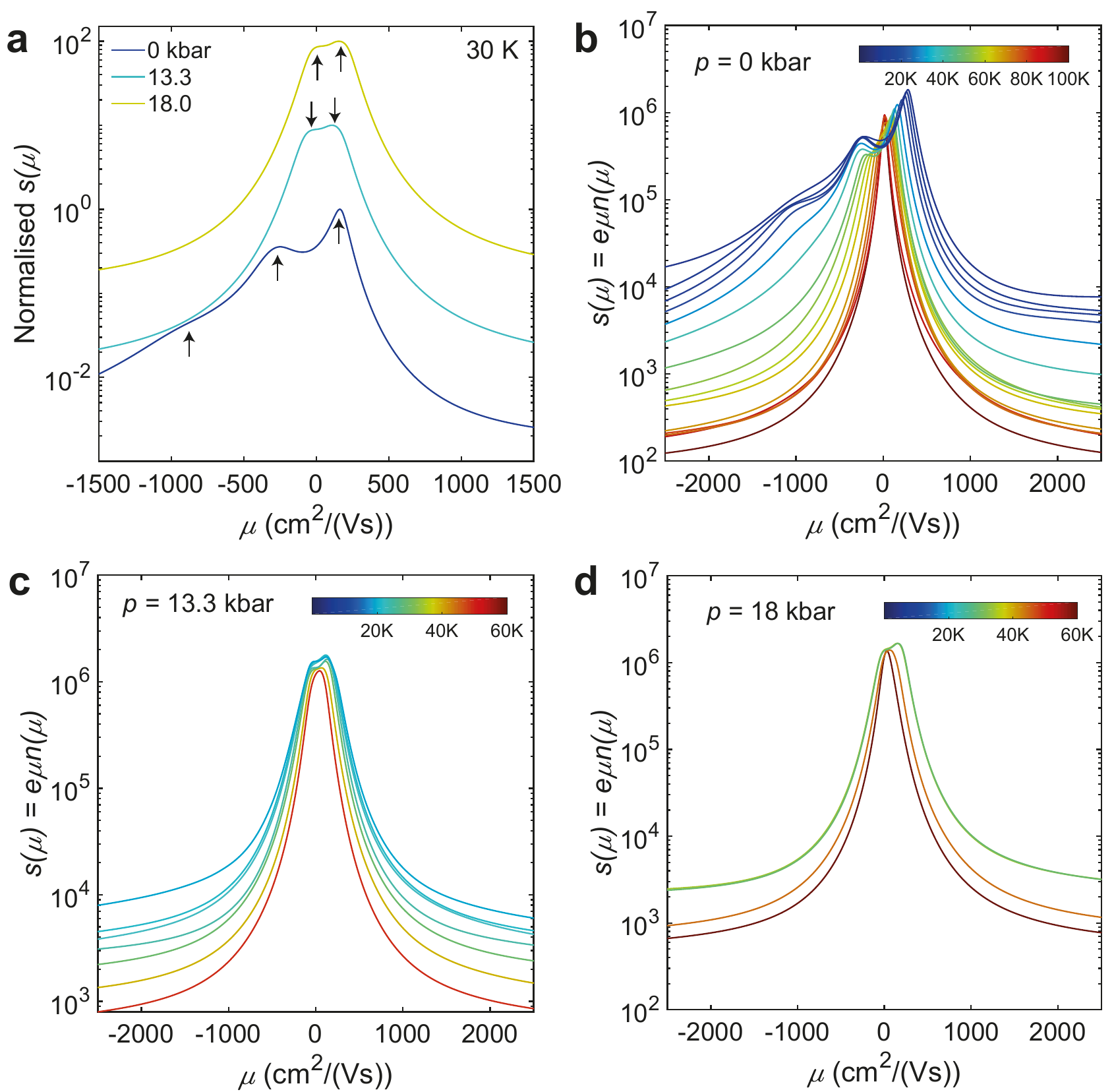}
	\caption{{\bf Mobility spectrum at different pressures.}
	(a) The mobility spectrum extracted from magnetotransport data \cite{Humphries2022}
for different pressures at 30\,K. Arrows indicated the positions of mobilities for the positive and negative charge carriers.
	Mobility spectrum extracted for constant temperatures at different pressures
of (b) 0\,kbar, (c) 13.3\,kbar and (d) 18\,kbar, respectively.}
	\label{SMFig_mobilitySpectrums}
\end{figure*}

\begin{figure*}[htbp]
	\centering
	\includegraphics[trim={0cm 0cm 0cm 0cm}, width=1\linewidth,clip=true]{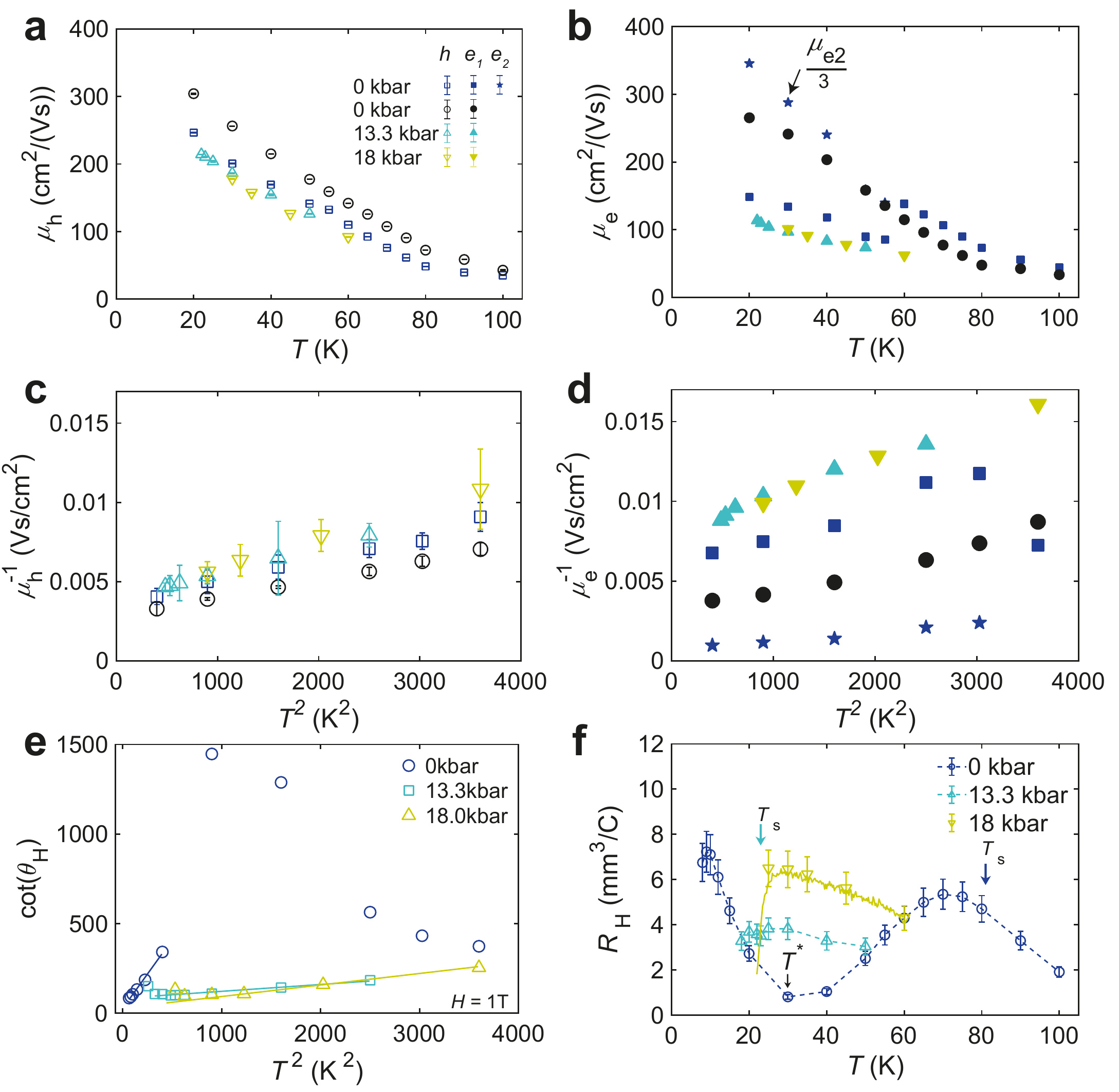}
	\caption{{\bf Mobility spectrum at different pressures.}
	The mobility temperature dependence for (a) positive charge carriers
(holes) and (b) the negative charge carriers (electrons).
The ambient pressure data  is modelled using a there-band model, described in the Ref.~\onlinecite{Zajicek2022},
 whilst the two-band model is taken from low-field limit ($<7$~T) only.
 The mobility of the second electron carrier, e$_{\rm 2}$, in (b) is scaled by a factor 1/3.
	The inverse mobility against a $T^{2}$ dependence for (c) holes and (d) electrons.
	(e) cot($\theta_{\rm H}$), cotangent of the Hall angle defined, temperature dependence.
Solid lines are fit to linear regions at higher temperatures. The values are taken from field-dependent measurements and chosen at a fixed field value for all temperatures.
	(f) The Hall coefficient, $R_{\rm H}$, temperature dependence for different pressures of Cu-FeSe. The dashed lines are guide to the eye.
 The solid line for 13.3\,kbar is from a temperature sweep measured in constant field of 1~T.
 $T^{*}$ marks the minimum in $R_{\rm H}$ for 0\,kbar.}
	\label{SMFig_MobilityExtras}
\end{figure*}

\end{document}